\documentclass[conference,10pt,letterpaper]{IEEEtran}
\IEEEoverridecommandlockouts
\usepackage{cite}
\usepackage{amsmath,amssymb,amsfonts}
\usepackage{algorithmic}
\usepackage{graphicx}
\usepackage{textcomp}
\usepackage{xcolor}
\usepackage{booktabs}
\usepackage{tabularx}
\usepackage{hyperref}
\usepackage{fancyhdr} 
\usepackage{setspace} 

\def\BibTeX{{\rm B\kern-.05em{\sc i\kern-.025em b}\kern-.08em
    T\kern-.1667em\lower.7ex\hbox{E}\kern-.125emX}}
\begin{document}

\title{Prerequisite Structure Discovery In Intelligent Tutoring Systems}


\author{Louis Annabi$^{1}$ and Sao Mai Nguyen$^{2}$

\thanks{This work was  supported by Inter Carnot MINES - TSN}

\thanks{$^{1}$ Flowers Team, U2IS, ENSTA Paris, Institut Polytechnique de Paris \& Inria, Palaiseau, France
        {\tt\small louis.annabi@gmail.com}}

\thanks{$^{2}$ Flowers Team, U2IS, ENSTA Paris, Institut Polytechnique de Paris \& Inria, Palaiseau, France
        {\tt\small nguyensmai@gmail.com}}%
}

\maketitle
\thispagestyle{fancy}
\lhead{}
\chead{
\texttt{
\begin{spacing}{0.9}
\scriptsize{
Annabi and Nguyen. Prerequisite structure discovery for an intelligent tutoring system based on intrinsic motivation. IEEE ICDL. 2023. \url{https://doi.org/10.1109/ICDL55364.2023.10364416}
 }
 \end{spacing}
}
\vspace{20pt}}
\rhead{}
\cfoot{}

\begin{abstract}
This paper addresses the importance of Knowledge Structure (KS) and Knowledge Tracing (KT) in improving the recommendation of educational content in intelligent tutoring systems. The KS represents the relations between different Knowledge Components (KCs), while KT predicts a learner's success based on her past history. The contribution of this research includes proposing a KT model that incorporates the KS as a learnable parameter, enabling the discovery of the underlying KS from learner trajectories. The quality of the uncovered KS is assessed by using it to recommend content and evaluating the recommendation algorithm with simulated students.
\end{abstract}

\begin{IEEEkeywords}
    Intelligent Tutoring Systems, Knowledge Tracing, Knowledge Structure Discovery
\end{IEEEkeywords}

\section{Introduction}

\subsection{Context}

In recent years, online education platforms have gained immense popularity, leading to a growing interest in automated methods for recommending pedagogical content.  Intelligent Tutoring Systems (ITS) are computer systems that provide personalised recommendation of educational content to optimise learners' progress, for instance  in the form of social robots for education \cite{Belpaeme2018SR} or massive open online courses.  

An ITS is generally decomposed into four components: (i) the \textit{domain knowledge} about the educational concepts, rules and problem solving strategies. It especially includes the list of skills or concepts that need to be mastered, called Knowledge Components (KCs), the list of KCs necessary to complete each exercise, as well as the relations between KCs, called the Knowledge Structure (KS); (ii) a \textit{student model} that estimates the evolution of the learner's knowledge states; (iii) a \textit{tutoring model} that recommends educational content (in our case exercises) to the learners, possibly based on the domain knowledge and on a student model; (iv) a user interface.

ITS often aim at maximising the learning progress of the student on the long run, thus relying on a model of the development of the learner's cognitive skills. Learning progress is both the natural extrinsic reward a tutor would try to maximise, and the intrinsic motivation for the learner, as theorised in developmental psychology \cite{Deci1975} and modeled for developmental systems \cite{Oudeyer2009FN}. Typically, pedagogical activities (in our case exercises) that maximise learning progress should be neither too hard nor too easy for the learner. This idea aligns with the concept of Zone of Proximal Development (ZPD) introduced by Vygotsty \cite{vygotsky1978mind}. The ZPD refers to the set of activities a learner is unable to do by herself but can do with some assistance. Because learning is most effective in the ZPD, several works have proposed implementations of this concept in tutoring models: 
\cite{Clement2015JEDMJ} propose a tutoring model based on the automatic estimation of the ZPD by exploiting domain knowledge given by experts, while \cite{mu2018combining} and \cite{vainas2019gotsky} use student models to estimate respectively student forgetting and student probability of success on given exercises.

The ZPD actually evolves as the learner acquires new knowledge. Knowledge Tracing (KT) corresponds to the task of accurately estimating the evolving knowledge states of learners on the KCs, based on their history of exercises and answers \cite{corbett1994knowledge}. KT models are typically trained as sequence prediction models, using recorded trajectories of students. As such, they jointly perform the task of tracing student knowledge and modeling student trajectories. For this reason, we indistinctively refer to KT models and student models. 






While machine learning techniques have been widely applied to KT and tutoring models, only few have focused on domain knowledge and in particular KS discovery, when most rely on an expert-given domain knowledge. Yet, knowing the KS can improve the accuracy of KT models \cite{chen2018prerequisite}, but most importantly can directly be exploited by a tutoring model in order to organize exercises into curricula \cite{Clement2015JEDMJ}.

\subsection{Problem and Contribution}

\begin{figure}[htb]
    \centering
    \includegraphics[width=0.48\textwidth]{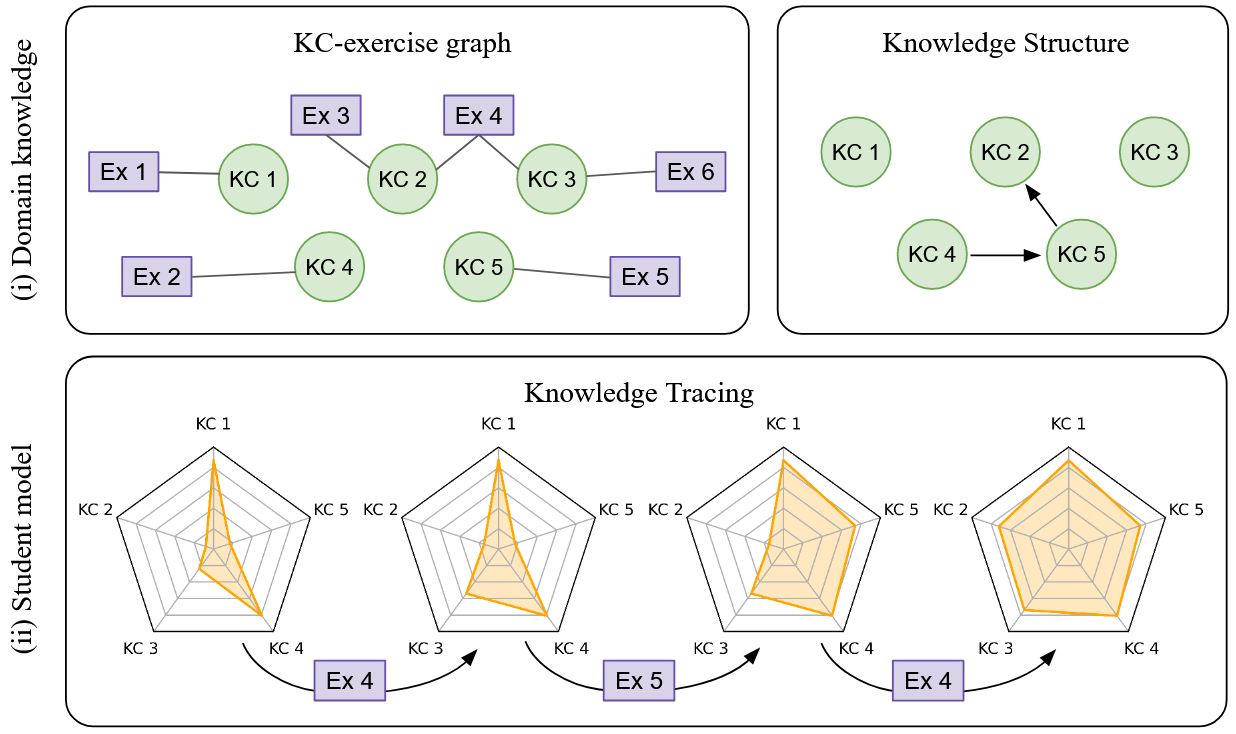}
    \vspace{-0.5em}
    \caption{Domain knowledge (top) and student models (bottom) can be exploited by tutoring models. The KC-exercise graph gives the KCs practiced on each exercise. The Knowledge Structure (KS) is represented as a directed acyclic graph where the edges represent prerequisite relations between KCs. Knowledge Tracing (KT) models predict the learner's level on each KC at each step of practice, here for a learner trajectory of length T=3. }
    \vspace{-1.5em}
    \label{fig:introduction_figure}
\end{figure}

 In this work, we depart from the common hypothesis of expert-given KS, and instead explore the use of machine learning techniques to discover prerequisite relations from recordings of learners interacting with an ITS. The KS is generally represented as a graph, where nodes are the KCs and edges are the relations between these KCs, as illustrated in Fig. \ref{fig:introduction_figure}. These relations represent for instance the prerequisite relations between KCs as in \cite{chen2018prerequisite}. 

 We propose a KS discovery method  using only  learners' performance data. Our method relies on a KT model conditioned on the KS: the KT model assumes that the learner's success on an exercise is conditional on her skill level on the exercise's KCs and prerequisite KCs. The prerequisite graph is learned together with the other parameters of the KT model, using backpropagation through time.

To exploit learners' performance data for deriving a KS, we can unfortunately not rely on publicly available datasets, where the exercise sequences are determined based on prior domain knowledge, which can be erroneous, incomplete or not available. 
The first reason is that the exercise sequencing is biased and data necessary to discover some prerequisite relations from the learners' performances might be missing in the designed curriculum. For instance, for an ITS for basic arithmetic concepts such as addition and multiplication, if exercises on multiplication are only recommended after exercises on addition, it is impossible to discover that there is a prerequisite relation between addition KC and multiplication KC. Indeed, we would need observations of attempts of exercises addressing the multiplication KC while still not proficient on the addition KC -- such observations could be absent from the dataset.  Second, the available datasets have generally been obtained with exercise sequences chosen by an expert or a tutoring model biased by expert knowledge, whereas we wish to address the problem even when the domain knowledge is unknown and no expert is available. 
Therefore, a relevant evaluation of KS discovery algorithms must use data with exercise sequences agnostic of the domain knowledge, to remove the possibility that the algorithm discovers the KS owing to this expert knowledge instilled in the exercise sequences, and to ensure that KS discovery methods capture information from user performances even in the absence of prior domain knowledge.
Finally, evaluation of the discovered KS requires comparing with the ground truth KS -- which is absent from publicly available datasets \cite{abdelrahman2023knowledge}. 

While building our own dataset would only ensure the validity of the results on a specific domain application, we chose in this work to use synthetic data generated by a general student model, as it also makes it possible to extensively experiment with different tutoring systems interacting with the student model. We experimentally verify in section \ref{sec:experiments} that exercise sequences agnostic of domain knowledge provide richer data in order to discover KS.

We generate learner trajectories using a complex student model incorporating KS, variations of exercise difficulty and learner profiles, and learner forgetting. 
We evaluate our KS discovery algorithm by: (1) measuring correctness of the KS recovered from learner trajectories compared to the ground truth KS, (2) a comparative evaluation with both a KS-based tutoring model exploiting the discovered graphs, and KT-based tutoring models. We use an adaptation of the ZPDES algorithm \cite{Clement2015JEDMJ} to recommend exercises based on the discovered KS. 

In summary, this article presents the following contributions:

\begin{itemize}
    \item We introduce a KT model that incorporates the KS as a learnable parameter, that we name Prerequisite Knowledge Tracing (PKT). With the given KS-exercise graph and by training PKT on learner trajectories data, we can uncover the underlying KS.

    \item We quantitatively evaluate our KS discovery method in comparison with other approaches from the literature.


    \item We qualitatively evaluate the uncovered KS as a basis for estimating and updating the ZPD \cite{vygotsky1978mind} using a tutoring model adapted from the ZPDES algorithm \cite{Clement2015JEDMJ}.
\end{itemize}

\section{Related Work}

\subsection{Intelligent Tutoring Systems}


Intelligent Tutoring Systems (ITS) can be considered as recommendation systems for educational content. The recommendation task has been formulated in the Partially Observable Markov Decision Process (POMDP) framework \cite{rafferty2016faster,clement2016comparison,shen2018improving}, as a Multi-Armed Bandit (MAB) problem \cite{Clement2015JEDMJ,segal2018combining}. We refer to \cite{Doroudi2019IJAIE} for a review of reinforcement learning based approaches for tutoring. 

While domain knowledge and student models can be difficult to acquire, they have the potential to significantly improve recommendation of pedagogical contents. For instance, \cite{clement2018adaptive} showed that tutoring models exploiting domain knowledge could yield better learner progress than expert sequences of pedagogical contents. We can distinguish two situations, where the tutoring model (i) exploits the KS (for instance \cite{clement2016comparison}) ; and (ii)  exploits a given or learned student model (for instance with POMDP planning \cite{rafferty2016faster}). Our approach belongs to the first category, using a KS estimated by our algorithm. 


\subsection{Knowledge Tracing}

KT corresponds to the task of accurately estimating the evolving knowledge states of learners on the different KCs, based on their history of exercises and answers. 
Bayesian Knowledge Tracing (BKT) \cite{corbett1994knowledge} pioneered this field of research by proposing a KT model that tracks students' knowledge states using hidden Markov models. The knowledge states are represented by a vector of binary values, each value indicating whether the corresponding KC is acquired by the learner. Other methods use continuous values \cite{baker2001basics} or vector representations \cite{tong2020structure} to model the proficiency on each KC. 

More recent KT research focuses on deep learning methods, pioneered by the Deep Knowledge Tracing (DKT) model \cite{piech2015deep}. DKT uses Long Short-Term Memory networks (LSTMs) \cite{hochreiter1997long} to model the temporal evolution of students' knowledge states encoded in the LSTM hidden states. Following the success of DKT, many deep learning architectures specialised on KT have been proposed,  using key-value memories \cite{zhang2017dynamic}, graph neural networks \cite{nakagawa2019graph,yang2021gikt} or Transformers \cite{shin2021saint+}. We refer to \cite{abdelrahman2023knowledge} for a recent and more complete review of the KT literature.

Some works have also studied ways to exploit given relations between KCs in order to improve the accuracy of KT models. The Graph Knowledge Tracing (GKT) model \cite{nakagawa2019graph} updates the knowledge states of the KCs of the completed exercise, as well as the knowledge states for parent and children KCs in a given relation graph. The Structure-based Knowledge Tracing (SKT) model \cite{tong2020structure} conditions the knowledge state updates on two distinct graphs. An undirected graph represents similarity relations between KCs, while a Directed Acyclic Graph (DAG) represents prerequisite relations between KCs. The model presented in \cite{chen2018prerequisite} exploits known prerequisite relations as soft constraints on the ordering of the estimated learner level on each KC, embedded in the loss function of the model. These models show that integrating knowledge on the KS can improve the accuracy of KT models. As such, improving KS discovery methods can lead to better KT.

\subsection{Knowledge Structure with Prerequisite Structure Learning}


As such, some works also propose strategies to uncover the prerequisite structure in case it is not provided by experts of the educational domain. Some methods \cite{piech2015deep, zhang2016topological} build heuristics in order to extract the implicit information about the KS from trained DKT models. In \cite{gan2022knowledge}, the authors compare different indices that can be used to induce relations between KCs based on learners data. Among the indices they experiment with, only the adjusted Kappa is asymmetric and can be used to build a DAG. \cite{scheines2014discovering} propose to frame KS discovery as a problem of causal structure discovery, that they solve using statistical independence tests, with simplifying assumptions.

In \cite{nakagawa2019graph}, several methods are proposed for KS discovery. A first method relies on observed statistics in the exercise sequences taken by the learners, based on the observation that the available data usually comes from online educational platforms where exercises are taken following an expert-given curriculum. 
If exercises are given randomly to learners, this method becomes unable to uncover the prerequisite relations. The second method proposes to learn the prerequisite relations, represented by an adjacency matrix, together with the other parameters of the neural network. However, because the graph in the GKT model \cite{nakagawa2019graph} is not constrained to represent prerequisite relations, it is not straightforward to induce the prerequisite structure from the learned graph. Similar to GKT, our KS discovery method adopts a KT-based approach, learning the prerequisite graph together with the other parameters of a KT model. We design the KT model so that the learnable graph explicitly corresponds to prerequisite relations, which allows us to outperform other KS discovery methods.


\section{Methods}

Here we describe the proposed method for prerequisite structure learning, as well as the tutoring model used to evaluate the discovered KS. We provide the code for all our implementations in a git repository\footnote{https://github.com/sino7/KS-discovery-for-ITS}.

\subsection{KS Discovery}

We describe Prerequisite Knowledge Tracing (PKT), a straightforward KT model conditioned on the prerequisite relations between the KCs. The prerequisite relations are represented by an adjacency matrix $M$ such that $M_{ij}=1$ if the $i$-th KC is a prerequisite for the $j$-th KC. With this adjacency matrix, the model predicts the probability of success of a student $s$ on an exercise $e$ as:

\vspace{-1em}
\begin{equation}
    p_{e, s, t} = p_g + (1 - p_s - p_g) \sigma \Big(softmin_{k \in \mathcal{P}_e} \big( \lambda_{s, k, t} \big) - \delta_{e} \Big)
\end{equation}

where $p_g$ and $p_s$ respectively denote the \textit{guess} and \textit{slip} probabilities, i.e. the probability of completing (resp. failing) the exercise without the required skills (resp. while mastering the required skills). $\sigma$ denotes the sigmoid function, $\lambda_{s, k, t}$ denotes skill level of learner $s$ on the $k$-th KC at time $t$, and $\delta_e$ denotes the exercise $e$ difficulty level. Finally, $\mathcal{P}_e$ denotes the set of KCs corresponding to the exercise $e$, and parent KCs in the prerequisite graph $M$. We can see that the estimated probability of success depends on the adjacency matrix $M$. Thanks to this \textit{softmin} aggregation, if the learner skill on one of the prerequisite KCs is insufficient, the probability of completing the exercise will be low. $\lambda_{s, k, t}$ is estimated as:

\begin{equation}
    \lambda_{s, k, t} = \mu_{s, k} + \alpha_s  S_{s, k, t} + \beta_s  F_{s, k, t}
\end{equation}

where $\mu_{s, k}$ denotes the initial skill level of the learner $s$ on the $k$-th KC, $S_{s, k, t}$ denotes the number of times the learner $s$ has completed with success an exercise related to the $k$-th KC before time $t$, and $F_{s, k, t}$ denotes the number of times the learner $s$ has failed an exercise related to the $k$-th KC before time $t$. $\alpha_s$ and $\beta_s$ are scalar parameters of the model, learned for each individual student $s$, that can be interpreted as the amount of skill gained respectively by succeeding and failing an exercise. 

The parameters \{$p_g, p_s, \delta_e, \mu_{s, k}, \alpha_s, \beta_s$, $M$\} are optimised in order to minimise the binary cross entropy loss between the predicted probability of success $p_{e, s, t}$ and the ground truth learner response. We use L2 regularisation on the parameters $\alpha_s$, $\beta_s$ and $\mu_{s, k}$ to prevent overfitting. Finally, we also regularise the adjacency matrix using the L1 norm to encourage sparse prerequisite graphs.

The proposed model implements the idea that a learner does not successfully complete exercises related to a certain KC if she has not first mastered the prerequisite KCs.









\subsection{Tutoring model}
\label{sec:tutoring_models}

\begin{figure}[htb]
    \centering
    \includegraphics[width=0.48\textwidth]{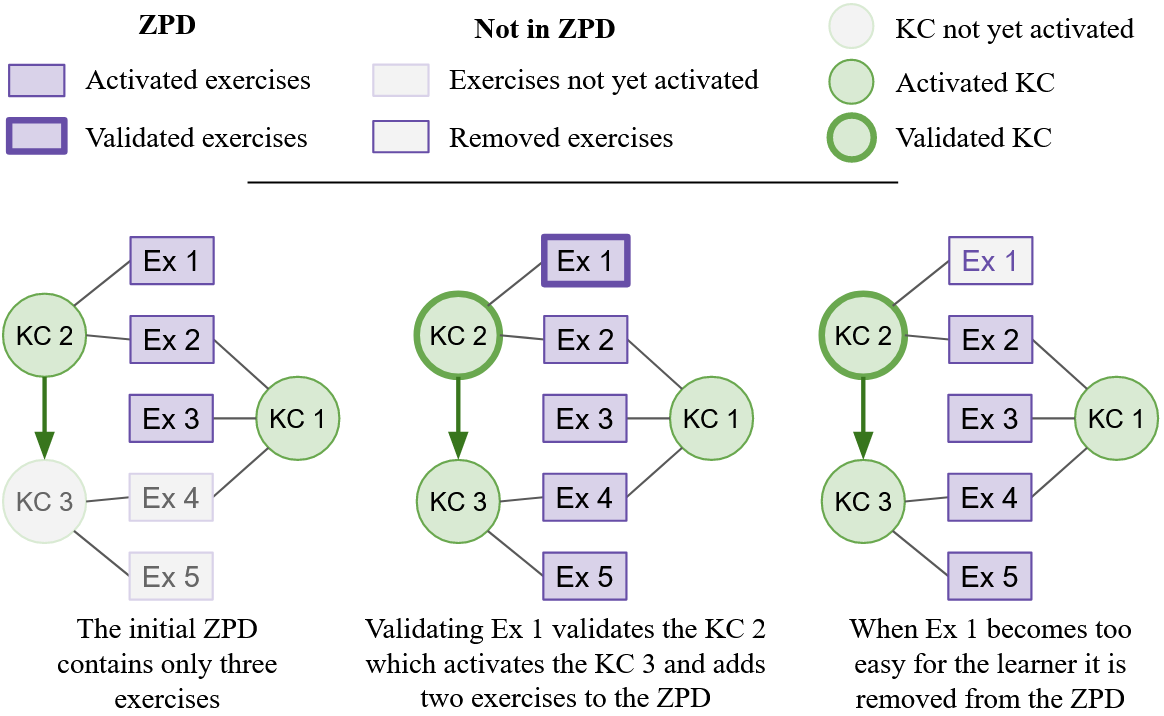}
    \vspace{-0.6em}
    \caption{Illustration of the ZPDES-KS method applied on a toy example. Learner progress can unlock new KCs and add new exercises to the ZPD, and can also deactivate exercises when success becomes too frequent.}
    \vspace{-1.3em}
    \label{fig:zpdes_ks_figure}
\end{figure}

The tutoring model we present is an adaptation of the ZPDES algorithm \cite{Clement2015JEDMJ}, that we label ZPDES-KS. 
ZPDES uses as input a progression graph on exercises to structure the curricula of learners. It builds and updates a personalised pool of exercises for each learner, representing their individual ZPD. As such, this ZPD should always contain exercises on which the learner has the potential to progress, i.e. exercises that are neither too easy nor too difficult. We adapt ZPDES to use a prerequisite graph on KCs. An illustration of the ZPDES-KS algorithm on a toy example is provided in Fig. \ref{fig:zpdes_ks_figure}. The ZPD expands with new exercises when new KCs are unlocked, and  exercises are removed when they become too easy. We  propose the following rules to build and update the ZPD:

\begin{enumerate}
    \item The ZPD is initialised with exercises for which all the KCs have no prerequisite in the prerequisite graph.
    \item If the empirical success on one exercise reaches a threshold value $\epsilon_v$, we consider the exercise to be validated.
    \item A KC is validated if it has at least one validated exercise.
    \item A KC is activated if all its prerequisite (parent) KCs are validated.
    \item An exercise is added if all its KCs are activated.
    \item If the empirical success on one exercise reaches a threshold value $\epsilon_r$, it is removed from the ZPD.
\end{enumerate}

The empirical success level $\hat{S}_{s, e}$ of the learner $s$ on the exercise $e$ is initialised at zero, and updated whenever the exercise $e$ is taken by the learner, using the update rule:

\vspace{-0.5em}
\begin{equation}
    \hat{S}_{s, e} \gets (1-\theta_S) \hat{S}_{s, e} + \theta_S S_{s, t}
\end{equation}

where $S_{s, t} = 1 $ if the learner successfully completes the exercise, and $S_{s, t} = 0$ otherwise; and $\theta_S$ is an update rate.

As in ZPDES, exercises are recommended to the learner based on a multi-armed bandit algorithm, using as reward function the empirical progress $\hat{P}_{s, e}$ of the learner $s$ on the exercise $e$. $\hat{P}_{s, e}$ is initialised at 0, and updated with:

\vspace{-0.5em}
\begin{equation}
    \hat{P}_{s, e} \gets (1-\theta_P) \hat{P}_{s, e} + \theta_P (S_{s, t} - \hat{S}_{s, e})
\end{equation}

where $\theta_P$ is an update rate, and $(S_{s, t} - \hat{S}_{s, e})$ can be interpreted as a noisy measurement of learning progress. For instance, if $\hat{S}_{s, e}$ is already close to 1, then succeeding on the exercise ($S_{s, t} = 1$) should not convey a large reward.

As focusing teaching on activities that provide more learning progress can act as a motivational mechanism \cite{Oudeyer2016PBR}, we encourage the recommendation of exercises within the ZPD by adding a reward $r_{ZPD}$ to those exercises. The recommendation algorithm then samples the exercise to recommend using a softmax distribution based on this reward function. 






\section{Experiments}
\label{sec:experiments}

We evaluate our approach in comparison with other KS discovery methods. First, KS discovery methods are evaluated with regard to their ability to properly recover the ground truth KS. Second, all methods are evaluated on their ability to be exploited by the ZPDES-KS tutoring model.

\subsection{Synthetic data generation}
\label{sec:synthetic_data}

\begin{figure}[htbp]
    \centering
    \vspace{-3em}
    \includegraphics[width=0.45\textwidth]{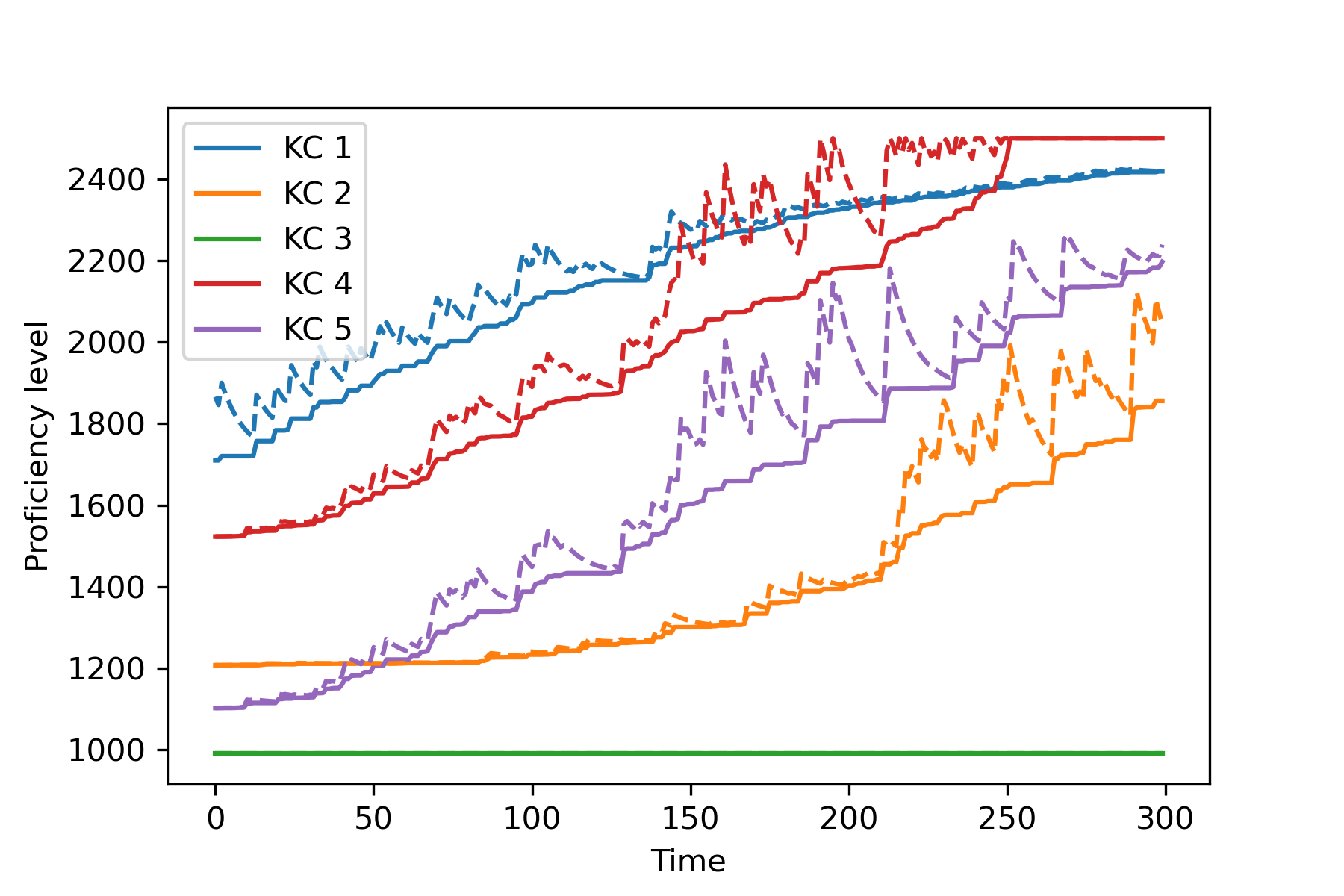}
    \vspace{-1.2em}
    \caption{Example of simulated learner trajectory. KC 3 is never practiced.}
    \vspace{-0.2em}
    \label{fig:example_ks}
\end{figure}

The student model used to generate synthetic data implements several mechanisms to account for prerequisite structure, variations of exercise difficulty and learner profiles, and learner forgetting. More details on our git repository$^1$. 

Learner forgetting is implemented by representing the learner proficiency both on the long-term and short-term. While short-term proficiency increases quickly when the KC is practiced, it is also forgotten at an exponential rate when the KC is not practiced, as proposed as forgetting curve in \cite{Wozniak1995ANE}. Concurrently, progressing on the long-term proficiency requires spaced practice of the KC. This is ensured by dividing the long-term learning by a factor corresponding to the gap between short-term and long-term proficiency. If the short-term proficiency is not given time to decrease back to the long-term proficiency level, this hinders long-term learning.

The prerequisite relation graph (KS), and KC-exercise graph are both randomly sampled. The KS is sampled using the Erdős–Rényi model with probability of edge $p=2/K$ where $K$ is the number of KCs. We only keep the upper triangular edges to make the graph directed, and then randomly shuffle the nodes. Finally, we also clean the graph of possible shortcuts. For instance if we have $i \to j \to k$ as well as $i \to k$, the prerequisite relation $i \to k$ is useless and we remove it. The KC-exercise graph is built in order to ensure that each exercise has at least one related KC, that each KC has at least one related exercise, and that an exercise cannot be related to two KCs if there exists a path in the KS connecting the two. We generate 10 random simulators using this model, with 10 KCs and 30 exercises. All the simulators use different KS and KC-exercise graphs. We sample 400 learner profiles and generate learner trajectories of length T=300. 

Fig. \ref{fig:example_ks} represents an example trajectory simulated with our student model, implementing the KS represented in Fig. \ref{fig:introduction_figure}.  Dashed lines represent the short-term proficiency and full lines the long-term proficiency. We observe that the learner only improves on the KC 5 once she is proficient enough on the KC 4, and on KC 2 once she is proficient enough on the KC 5. 
We also see that significant increases in long-term proficiency only occur once the short-term effect of previous practice vanishes.


\subsection{Prerequisite Structure Learning}

We experiment with different methods for finding the KS from the learners trajectories:
\begin{itemize}
    \item \textbf{PKT}: Using the proposed PKT model.
    \item \textbf{KI} : Kappa Index as proposed in \cite{gan2022knowledge} 
    \item \textbf{DKT-H} : Using the heuristic method proposed in \cite{zhang2016topological} on top of a trained DKT model \cite{piech2015deep}
    \item \textbf{SKT-H} : Using the heuristic method proposed in \cite{zhang2016topological} on top of a trained SKT model \cite{tong2020structure}
    \item \textbf{SKT-L} : Learning the adjacency matrix during the training of the SKT model \cite{tong2020structure}
\end{itemize}

Note that for the SKT model, we use the partial model representing only directed relations between the KCs, and not the full model that also implements undirected relations, since our primary goal is to discover the prerequisite relations.

Each method provides a matrix $M$ where $M_{ij}$ is the estimated strength of the prerequisite relation $i \to j$. We clean all the matrices to remove potential cycles, by ordering all the values in increasing order, and setting to 0 the values $M_{ij}$ that participate in cycles in the graph. After this step, we find for each method a threshold value $\theta$ maximizing the average f1-score between the adjacency matrix $A$ defined by $\big\{A_{ij} = (M_{ij}>\theta), \forall{ij}\big\}$ and the ground truth graph $A^*$.

With all 10 synthetic data generators, we sample 400 learner profiles and generate learner trajectories of length T=300, with two different scenarios. In the first scenario, exercises provided to the learners are sampled randomly. In the second scenario, exercises sequences follow a predetermined order designed using half of the prerequisite relations of the ground truth KS.

\begin{table}[htbp]
\vspace{-0.3cm}
  \centering
  \caption{Evaluation of the KS discovery methods. We report f1-scores for each method, with data collected (i) with random exercise sequences, and (ii) with partial information of the KS.}
\vspace{-0.3cm}
  \begin{tabularx}{0.48\textwidth}{@{} *{7}{X} @{}}
    \toprule
      & & PKT & KI & DKT-H & SKT-H & SKT-L \\
    \midrule
    \multicolumn{2}{l}{Random seq.} & \textbf{0.46} & 0.27 & 0.19 & 0.22 & 0.37 \\
    \multicolumn{2}{l}{Informed seq.} & 0.17 & \textbf{0.23} & 0.18 & \textbf{0.23} & 0.17 \\
    \bottomrule
  \end{tabularx}
  \label{tab:ks_discovery_results}
\vspace{-0.4cm}
\end{table}

Table \ref{tab:ks_discovery_results} reports the f1-scores for each method. We use binary classification metrics for evaluation, where the two classes correspond to the presence or absence of a directed edge in the graph. 
We can observe that the two methods that learn the adjacency matrix with a KT model perform better,  PKT outperforming the more complex SKT model. 
Besides, we observe that using random exercise sequences leads to better scores. This validates the first argument given in introduction to justify our choice of not using publicly available datasets: the biased sequencing of exercises in those datasets can hinder KS discovery.


\vspace{-0.3em}
\subsection{Exploiting prerequisite relations for recommendation}

\begin{figure}
    \centering
    \includegraphics[width=0.48\textwidth]{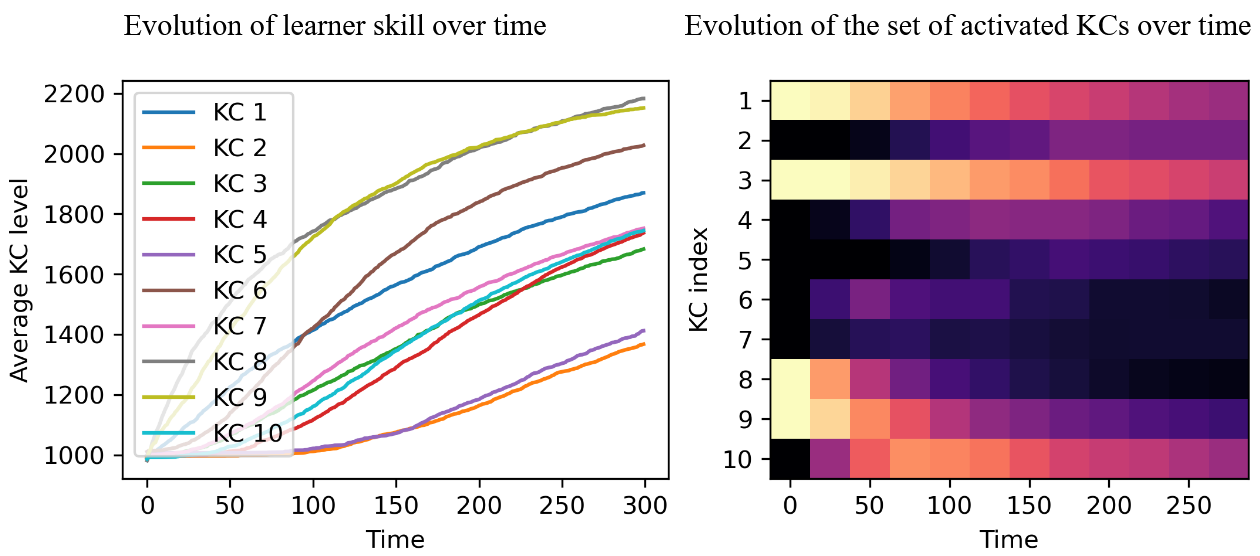}
\vspace{-0.3cm}
    \caption{Results with the ZPDES-KS algorithm on one of the datasets, with the KS discovered by the PKT method. Left: learner skills over time, averaged on 300 learners. Right: activated KCs in the ZPDES-KS algorithm over time, averaged on 300 learners (clear=activated).}
    \label{fig:example_zpdes}
    \vspace{-1.7em}
\end{figure}

Using the ZPDES-KS algorithm, we can build tutoring models from the discovered KS. We display in Fig. \ref{fig:example_zpdes} example results obtained with the ZPDES-KS tutoring model using the prerequisite relations found with the PKT method. Initially, only four the KCs are activated. The learners quickly progress on these KCs, which unlocks new KCs and adds new exercises in the ZPD. Once a threshold of empirical success level is reached, the KCs are deactivated.

\begin{table*}[htbp]
  \centering
  \caption{Evaluation of the tutoring models. Best values are in bold. Best values without taking into account the ZPDES-KS tutoring with ground truth prerequisite graph (GT) are underlined.}
  \vspace{-0.7em}
  \begin{tabularx}{\textwidth}{XX|X|XXXXXX|XXX}
    \toprule
    \multicolumn{2}{l|}{Tutoring model} & Random & \multicolumn{6}{c|}{ZPDES-KS} & \multicolumn{3}{c}{MBT} \\
    \midrule
    \multicolumn{2}{l|}{Graph / KT model} & & GT & PKT & KI & DKT-H & SKT-H & SKT-L & PKT & DKT & SKT \\
    \midrule
    \multicolumn{2}{l|}{Average level} & 1414 & \textbf{1629} & \underline{1528} & 1480 & 1402 & 1406 & 1450 & 1412 & 1411 & 1425 \\ 
    \multicolumn{2}{l|}{Final level} & 1737 & \textbf{2034} & \underline{1894} & 1826 & 1713 & 1727 & 1788 & 1738 & 1700 & 1796 \\ 
    \bottomrule
  \end{tabularx}
  \label{tab:tutoring_results}
  \vspace{-1.9em}
\end{table*}

We evaluate all the KS discovery methods according to their corresponding tutoring system, by measuring the average proficiency on all the KCs for 300 simulated learners practicing for T=300 iterations. The learners start with an average proficiency level of 1000 and progress at different paces according to the taken exercises. We report in table \ref{tab:tutoring_results} the average proficiency level throughout practice, and the average proficiency level at the end of practice. We also provide the results obtained when using the ground truth KS (labeled GT) with ZPDES-KS, as well as with random exercise sequences.

To further validate our methods, we compare the different KS-based tutoring models with a KT-based tutoring model, that we call Model-Based Tutoring (MBT). This tutoring model aims at maximising the expected progress of the learner, estimated using a given KT model. At each time step, for each exercise, the algorithm simulates the evolution of the knowledge state of the learner, and estimates the resulting progress on each KC. For each exercise, a score is computed as the expected progress averaged on the KCs. We then sample the exercise to recommend using a softmax distribution computed on these scores.
We observe that given the ground truth KS, ZPDES-KS outperforms all other approaches, despite not exploiting student models. Among the methods that automatically discover the KS, the proposed PKT approach significantly outperforms the baselines, while still remaining sub-optimal compared to the ground truth graph. Except for the SKT, the KT models do not provide better tutoring than random recommendations.

Overall, these results seem to indicate that domain knowledge in the form of prerequisite relations can be well discovered and exploited by tutoring models implementing the concept of ZPD. Interestingly, we obtain better tutoring strategies using only prerequisite relations than with trained student models, that could in principle encode more information such as exercise difficulty and student profiles. 

\section{Conclusion}

We have investigated the use of machine learning techniques in order to recover KS from student trajectories, without relying on domain knowledge given by experts. The proposed KS discovery method is based on a straightforward KT model conditioned on the KS, that we call PKT. The KS is learned together with the other parameters of the PKT model. The uncovered prerequisite relations can be used to deduce rules to estimate the ZPD of a learner interacting with the ITS. By recommending exercises within this ZPD, the tutoring model can maximise learning progress while keeping the learner engaged in her practice by stimulating her intrinsic motivation.

We have evaluated the accuracy of our KS discovery method, and shown that the PKT method outperforms other approaches from the literature. While the discovered graphs do not perfectly match the ground truth KS, we have also shown that tutoring models based on these graphs can perform better than tutoring models based on trained KT models.

Our results tend to show that KS-based recommendation can compete with KT-based approaches. They indicate the potential of more research on the topic of KS discovery, as this approach has been so far underexplored compared to the efforts invested in improving KT models. In future work, we would like to investigate the data efficiency of the KT models and KS discovery algorithms. Since accurate recommendation depends on the quality of the KT or KS,  we may have a dilemma between exploration and exploitation. While recommending random exercises might improve the KT model or the KS estimation (exploration), the learner would benefit from optimised exercise sequencing to maximise her progress (exploitation). 
Besides, while the ZPDES-KS algorithm accounts for prerequisite relations, we would like to improve it in order to also account for cognitive models of forgetting, taking inspiration from \cite{mu2018combining}.

\vspace{-1em}


\bibliographystyle{IEEEtran}

\bibliography{references}  

%
%
%
%
\end{document}